%% file: main.tex
\documentclass[conference]{IEEEtran}
\IEEEoverridecommandlockouts
\usepackage{booktabs}
\usepackage{multirow}
\usepackage{textcomp}
\usepackage{xcolor}
\usepackage{cite}
\usepackage{latexsym}
\usepackage{graphicx}
\usepackage{subcaption}
\usepackage{amsfonts,amssymb,amsmath}
\usepackage{nccmath}
\usepackage{optidef}
\usepackage{hyperref}
\usepackage{color,soul}
\usepackage{array}
\usepackage{algorithm}
\usepackage{algorithmic}
\usepackage[T1]{fontenc}
\usepackage[utf8]{inputenc}
\usepackage{fancyhdr}
\usepackage{lastpage}
\usepackage{caption}
\usepackage{soul}
\usepackage{subcaption}
\usepackage{setspace}
\usepackage{bm}
\def\BibTeX{{\rm B\kern-.05em{\sc i\kern-.025em b}\kern-.08em
    T\kern-.1667em\lower.7ex\hbox{E}\kern-.125emX}}
\usepackage{fancyhdr}
\fancypagestyle{firststyle}{
	\fancyhf{}
	\fancyhead[L]{H. Poddar, A. Chowdary, T. S. Rappaport, and M. Chafii, ``Full-Stack End-To-End Sub-THz Simulations at 140 GHz using NYUSIM Channel Model in ns-3'', \textit{in IEEE Wireless Communications and Networking Conference (WCNC) 2024}, Dubai, United Arab Emirates, April 2024, pp. 1--6.}

}

\begin{document}
\title{Full-Stack End-To-End Sub-THz Simulations at 140 GHz using NYUSIM Channel Model in ns-3 
\thanks{ This work is
supported by the NYU WIRELESS industrial affiliates program.}
}
\author{\IEEEauthorblockN{Hitesh Poddar$^{\dagger}$, Akhileswar Chowdary$^{\dagger}$, Theodore S. Rappaport$^{\dagger}$, Marwa Chafii$^{*\dagger}$}
\IEEEauthorblockA{$^{\dagger}$NYU WIRELESS, NYU Tandon School of Engineering, Brooklyn, NY, USA, \{hiteshp, akhileswar.chowdary, tsr\}@nyu.edu\\$^{*}$Engineering division, New York University Abu Dhabi, UAE, marwa.chafii@nyu.edu}
}
\maketitle

\thispagestyle{firststyle}

\begin{abstract}
The next generation of wireless communication is expected to harness the potential of the sub-THz bands to achieve exceptional performance and ubiquitous connectivity. However, network simulators such as ns-3 currently lack support for channel models above 100 GHz. This limits the ability of
researchers to study, design, and evaluate systems operating above 100 GHz. Here, we use the drop-based NYUSIM channel model to simulate channels above 100 GHz in all 3GPP scenarios including urban microcell (UMi), urban macrocell (UMa), rural macrocell (RMa), indoor hotspot (InH),
and indoor factory (InF). We evaluate the full stack downlink end-to-end performance (throughput, latency, and packet drop) experienced by a single user equipment (UE) connected to a Next Generation Node B (gNB) operating in the sub-THz bands for three gNB--UE antenna configurations: 8x8--4x4, 16x16--4x4, and 64x64--8x8 by using the NYUSIM channel model at 140 GHz in the ns-3 mmWave module. \textit{Our simulations demonstrate that sub-THz bands can enable high-fidelity applications that require data rates exceeding 1 Gbps and latency below 15 milliseconds (ms) using the current mmWave protocol stack, and large antenna arrays}. In addition, we show the variation in throughput vs number of realizations and find the optimal number of realizations required to obtain statistically significant results. We strongly encourage researchers worldwide to adopt a similar approach, as it enables the readers to assess the accuracy and reliability of the reported results and enhance the findings' overall interpretability.
\end{abstract}

\section{Introduction}
\input{Sections/Introduction.tex}

\section{Simulation Setup}\label{sec:simsetup}
\input{Sections/Simulation_setup.tex}

\section{Results and Discussions}\label{sec:results}
\input{Sections/Results.tex}
\subsection{Trade-off between latency, throughput and packet drop}\label{sec:tradeoff}
\input{Sections/Tradeoff.tex}
\subsection{Impact of the Number of Realizations on End-to-End Performance Metrics}\label{sec:confidenceinterval}
\input{Sections/Impact.tex}

\section{Conclusion}\label{sec:conclusion}
\input{Sections/Conclusion.tex}

\bibliographystyle{IEEEtran}
\bibliography{references}
\vspace{12pt}
\color{red}
\end{document}

%% file: Sections/Introduction.tex
6G communication holds tremendous potential to enable a wide array of applications, including autonomous navigation, smart cities \cite{viswanathan:2020:communications-in-the, aazhang:2019:key-drivers}, augmented and virtual reality (AR/VR) \cite{dang:2020:what-should}, haptics, integrated sensing and communication \cite{bazzi:2023:on-integrated-sensing}, and Industry 4.0 \cite{viswanathan:2020:communications-in-the, aazhang:2019:key-drivers}. To fully realize the capabilities of these cutting-edge applications, exceptional data rates, imperceptible latency, and ubiquitous connectivity are essential \cite{rappaport:2019:wireless-communications-and-applications,chaffi:2023:twelve-scientific-challenges}. However, the current mmWave bands, operating in the frequency range of 24--72 GHz, cannot meet the necessary performance requirements for these demanding applications due to limited bandwidth. Typically the maximum continuous bandwidth allocated in mmWave bands is limited to 400 MHz \cite{rappaport:2013:millimeter-wave-mobile, naga:2014:network-densification}. To overcome the bandwidth limitation, it is imperative to leverage the sub-THz bands, which span frequencies from 100--300 GHz and provide larger continuous bandwidths in the order of tens of GHz \cite{next-g-alliance:2022:next-g-alliance-report, rappaport:2019:wireless-communications-and-applications}. The Federal Communications Commission's (FCC) Office of Engineering and Technology, as outlined in the ET Docket No. 18--21 issued in 2019 \cite{fcc}, has taken proactive measures to foster the advancement of novel wireless technologies in the sub-THz bands. Furthermore, standardization bodies, industry, and academia have embarked upon an exploration of the D-band, centered around 140 GHz, as a promising candidate for 6G communication \cite{singh:2019:beyond-5g-THz-spectrum,next-g-alliance:2022:next-g-alliance-report}.
\par However, the official release of the 6G standard, encompassing the defining elements of the physical layer, MAC layer, and higher-layer procedures is scheduled to be released in 2026 \cite{chen:2023:5g-advacned-towards-6g}. As a result, this work adopts a pragmatic approach by leveraging the existing ns-3 mmWave module \cite{mezzavilla:2018:end-to-end-simulation-of-5g-mmwave} which implements the 5G NR protocol stack to evaluate the end-to-end performance of the 140 GHz (sub-THz) channel for all 3GPP-listed scenarios namely UMi, UMa, RMa, InH, and InF \cite{3gpp:2022:3gpp-ts-38.101-1,3gpp:2022:3gpp-ts-38.101-2}. Additionally, by systematically scaling up the antenna elements at both the gNB and UE, we aim to quantitatively determine the extent of performance enhancements achievable for the end-to-end performance metrics (throughput, latency, and packet drops). We conduct an extensive series of simulations using the NYUSIM channel model \cite{poddar:2023:ns-3-implementation-of-sub} at 140 GHz for all the 3GPP specified scenarios, implemented into the widely used ns-3 mmWave module \cite{mezzavilla:2018:end-to-end-simulation-of-5g-mmwave}.
\par NYUSIM is an active open-source mmWave and sub-THz channel simulator implemented in ns-3 \cite{poddar:2023:ns-3-implementation-of-sub,poddar:2023:full-stack,poddar:2023:nyusim-in-ns3} and MATLAB \cite{sun:2016:nyusim:-mmwave-and-sub-thz,sun:2017:a-novel-millimeter-wave-channel,ju:2019:a-millimeter-wave-channel-simulator,poddar:2023:a-tutorial-on-nyusim}. NYUSIM channel models are developed based on an exhaustive collection of field measurement data for the frequency range of 28 to 140 GHz\footnote{Although the measurements were at 142 GHz, we use 140 GHz interchangeably for simplicity. However, the simulations in this paper are at 140 GHz.}, conducted between 2011--2022. NYUSIM possesses the capability to generate channels for the frequency range of 0.5--150 GHz across diverse 3GPP-defined scenarios \cite{rappaport:2013:broadband-millimeter-wave-propagation, rappaport:2015:wideband-millimeter-wave-propagation, samimi:2016:3-d-millimeter-wave-statistical-channel, samimi:2016:local-multipath-model}.
\par This paper is structured as follows. In Section \ref{sec:simsetup}, we present the simulation setup. Section \ref{sec:results} presents results and insights. Furthermore, Subsection \ref{sec:tradeoff} illustrates the trade-off between latency, throughput, and packet drops. Additionally, Subsection \ref{sec:confidenceinterval} proposes an effective methodology for presenting statistical results derived from extensive wireless network simulations. Finally, in Section \ref{sec:conclusion}, we draw our conclusions.

%% file: Sections/Simulation_setup.tex
To thoroughly explore the full stack downlink end-to-end performance for a single UE connected to a gNB operating in the sub-THz band, we employ the NYUSIM channel model \cite{poddar:2023:ns-3-implementation-of-sub} in the widely-utilized ns-3 mmWave module \cite{mezzavilla:2018:end-to-end-simulation-of-5g-mmwave}. Our examination encompasses an evaluation of the end-to-end throughput, latency, and packet drop experienced by a wireless modem operating at 140 GHz with 1 GHz bandwidth (maximum bandwidth supported by NYUSIM \cite{sun:2016:nyusim:-mmwave-and-sub-thz,ju:2019:a-millimeter-wave-channel-simulator}) across various UMi, UMa, RMa, InH, and InF scenarios, accounting for both line-of-sight (LOS) and non-line-of-sight (NLOS) channel conditions.
\par The simulations consider three distinct gNB--UE antenna configurations, namely \textbf{\textit{Ant1: 8$\times$8 (gNB)--4$\times$4 (UE)}}, \textbf{\textit{Ant2: 16$\times$16 (gNB)--4$\times$4 (UE)}}, and \textbf{\textit{Ant3: 64$\times$64 (gNB)--8$\times$8 (UE)}}, respectively. The simulations focus on a single gNB and UE in fixed LOS/NLOS channel conditions, in a realistic channel generated by the NYUSIM channel model. A single UE and gNB are chosen because they facilitate a fundamental understanding of the system's peak achievable performance, devoid of intra or inter-cell interference that may arise in multi-user scenarios. The UE is fixed at a distance of 100 meters from the gNB, which is well within a typical small cell size of approximately 200 meters and for ease of comparison of performance across different scenarios at a fixed distance. The gNB transmits at a power level of 30 dBm. Video is transmitted from a remote server using user datagram protocol (UDP) at different source application rates ranging from 250 Mbps to 3000 Mbps \cite{qualcommVideo}. The choice of UDP for transmission stems from its lower overhead and accurate representation for link performance estimation than that of the transmission control protocol (TCP). TCP entails mechanisms for reliable data delivery, flow control, and congestion control, which introduces additional latency and processing overhead. In contrast, UDP operates as a connectionless protocol devoid of such mechanisms, thereby reducing overhead and minimizing latency. Furthermore, TCP's congestion control mechanisms can significantly impact the observed throughput, making it challenging to measure the actual link performance accurately.
\par We execute 2500 realizations (the reason for choosing this number is explained in Subsection \ref{sec:confidenceinterval}) for each channel condition (LOS/NLOS), scenario, antenna configuration, and source application rate. Within each realization, which spans a duration of 9 seconds, analog beamforming with a single spatial stream is employed, hybrid automatic repeat request (HARQ) processes are set to 8, and blockage effects are disabled (NYUSIM in ns-3 currently doesn't support blockage models) \cite{poddar:2023:ns-3-implementation-of-sub}. The size of the radio link control (RLC) buffer is set to 10 Megabytes. The remaining configuration parameters align with the default values specified within the mmWave module \cite{mezzavilla:2018:end-to-end-simulation-of-5g-mmwave}.

%% file: Sections/Results.tex
\begin{figure}[t!]
    \centering
    \includegraphics[width=9cm,height=10cm,keepaspectratio]{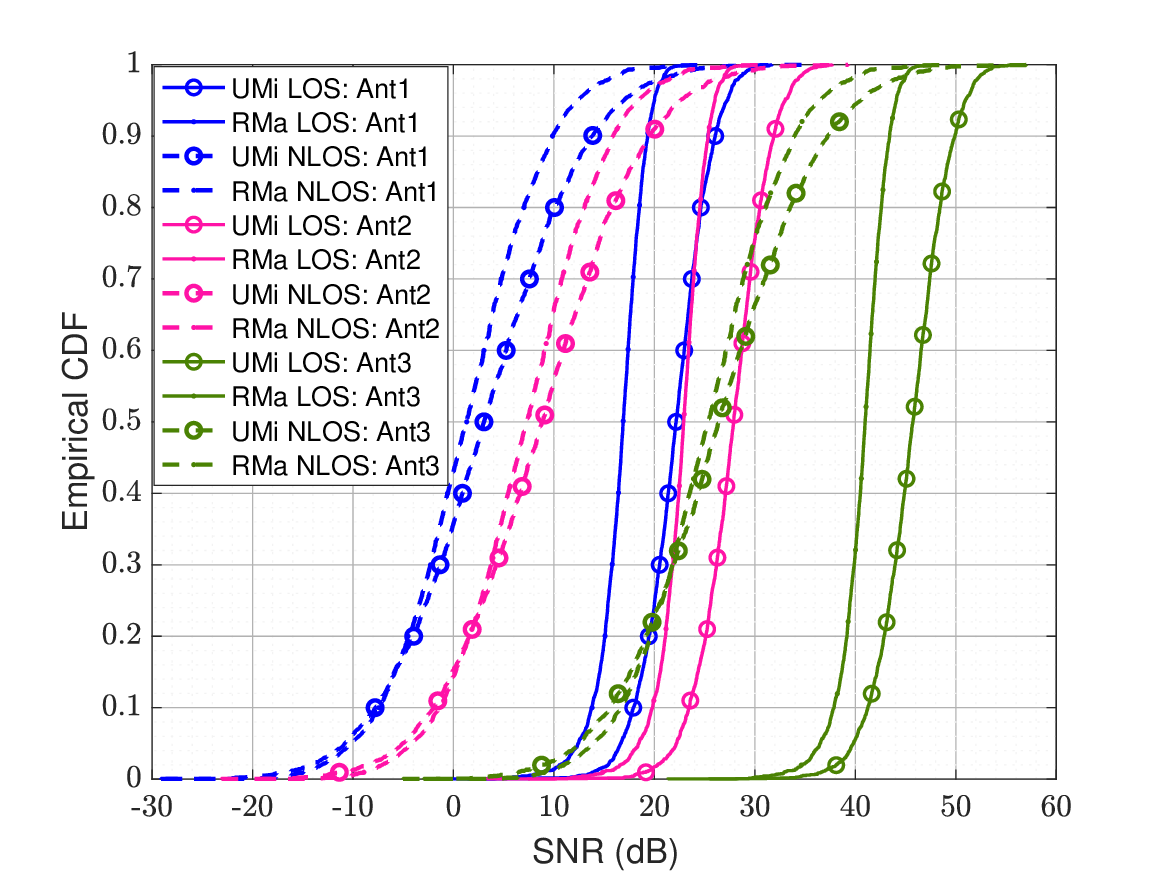}
    \caption{CDF of signal-to-noise ratio (SNR) at the UE for 3 antenna configurations of gNB and UE in Outdoor scenarios, namely UMi and RMa for 100 m gNB-UE separation distance, 30 dBm transmit power, 140 GHz frequency, 1 GHz bandwidth.}
    \vspace{-0.2in}
    \label{fig:outdoor_snr}
\end{figure}
\begin{figure}[t!]
    \centering
    \includegraphics[width=9cm,height=10cm,keepaspectratio]{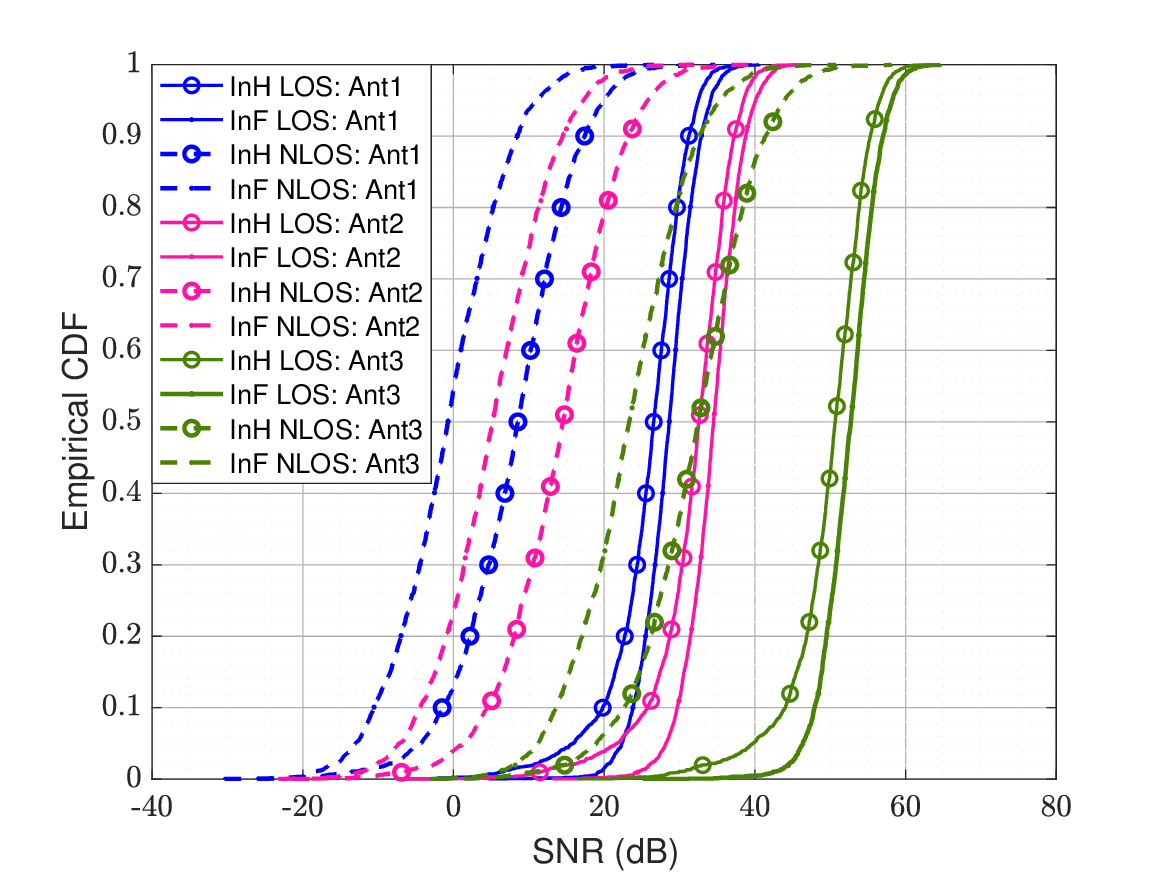}
    \caption{CDF of SNR at the UE for 3 antenna configurations of gNB and UE in Indoor scenarios (InH and InF) for 100 m gNB-UE separation distance, 30 dBm transmit power, 140 GHz frequency, 1 GHz bandwidth.}
    \vspace{-0.2in}
    \label{fig:indoor_snr}
\end{figure}
\begin{figure*}[htbp]
    \centering
    \begin{subfigure}[b]{0.2\textwidth}
        \centering
        \includegraphics[width=4.6cm,height=4.6cm,keepaspectratio]{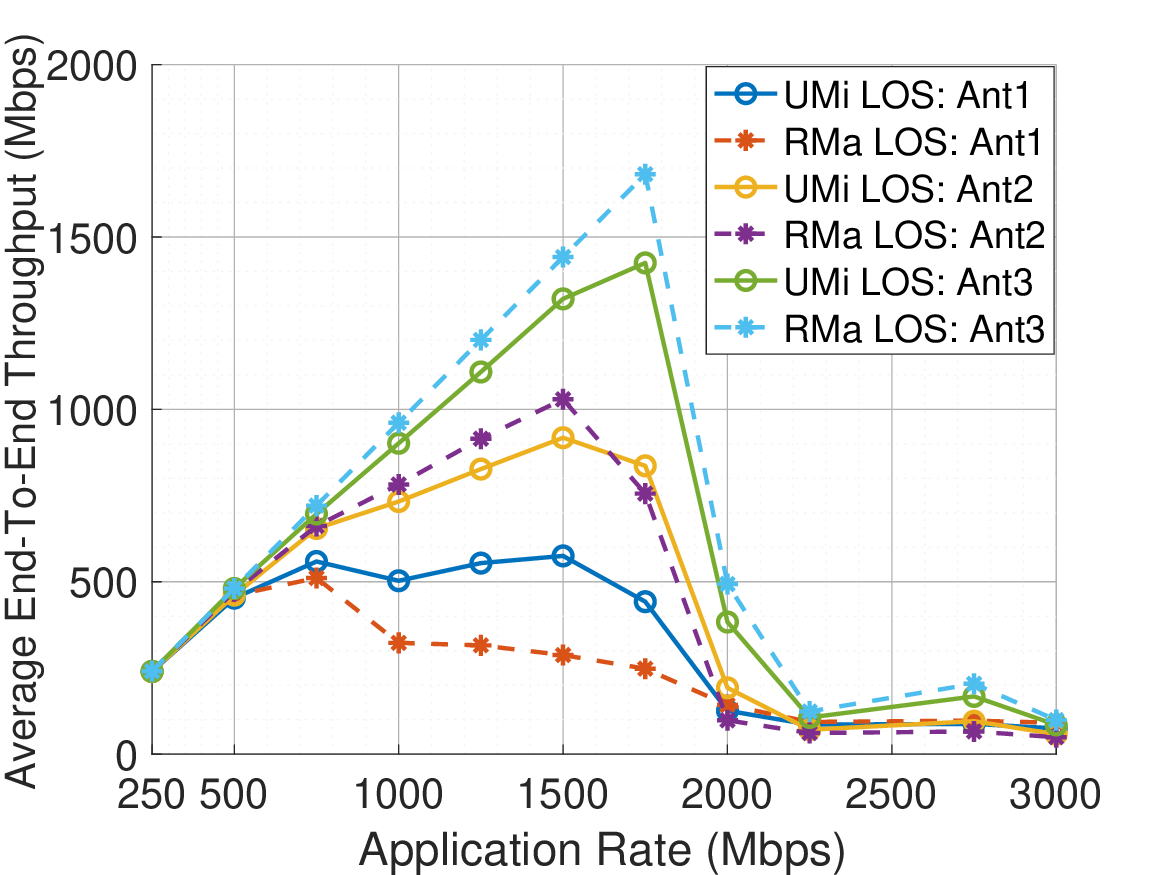}
        \caption{Average throughput vs application rate for 3 antenna configurations of gNB and UE in Outdoor (UMi and RMa) LOS scenarios.}
        \label{fig:outdoor_throughput_los}
    \end{subfigure}
    \hspace{0.2in}
    \begin{subfigure}[b]{0.2\textwidth}
        \centering
        \includegraphics[width=4.6cm,height=4.6cm,keepaspectratio]{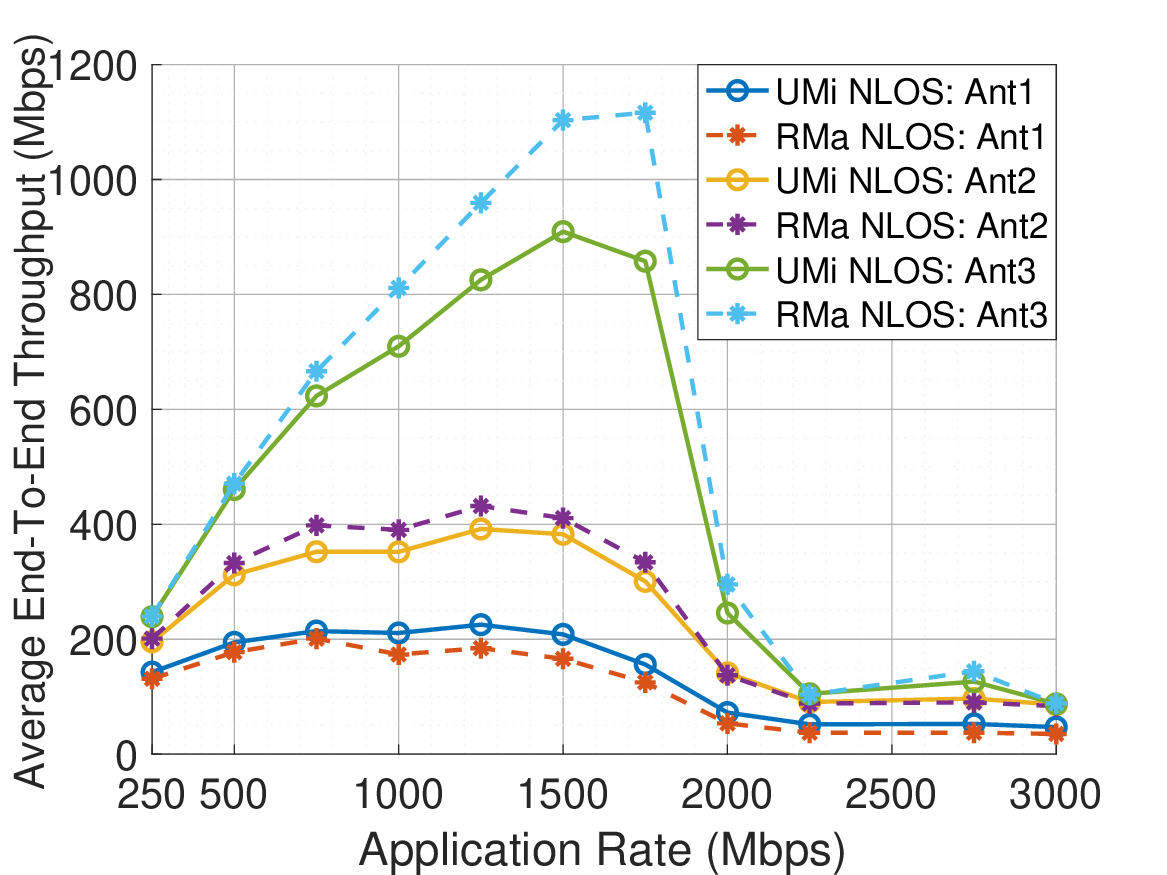}
        \caption{Average throughput vs application rate for 3 antenna configurations of gNB and UE in Outdoor (UMi and RMa) NLOS scenarios.}
        \label{fig:outdoor_throughput_nlos}
    \end{subfigure}
    \hspace{0.2in}
    \begin{subfigure}[b]{0.2\textwidth}
        \centering        \includegraphics[width=4.6cm,height=4.6cm,keepaspectratio]{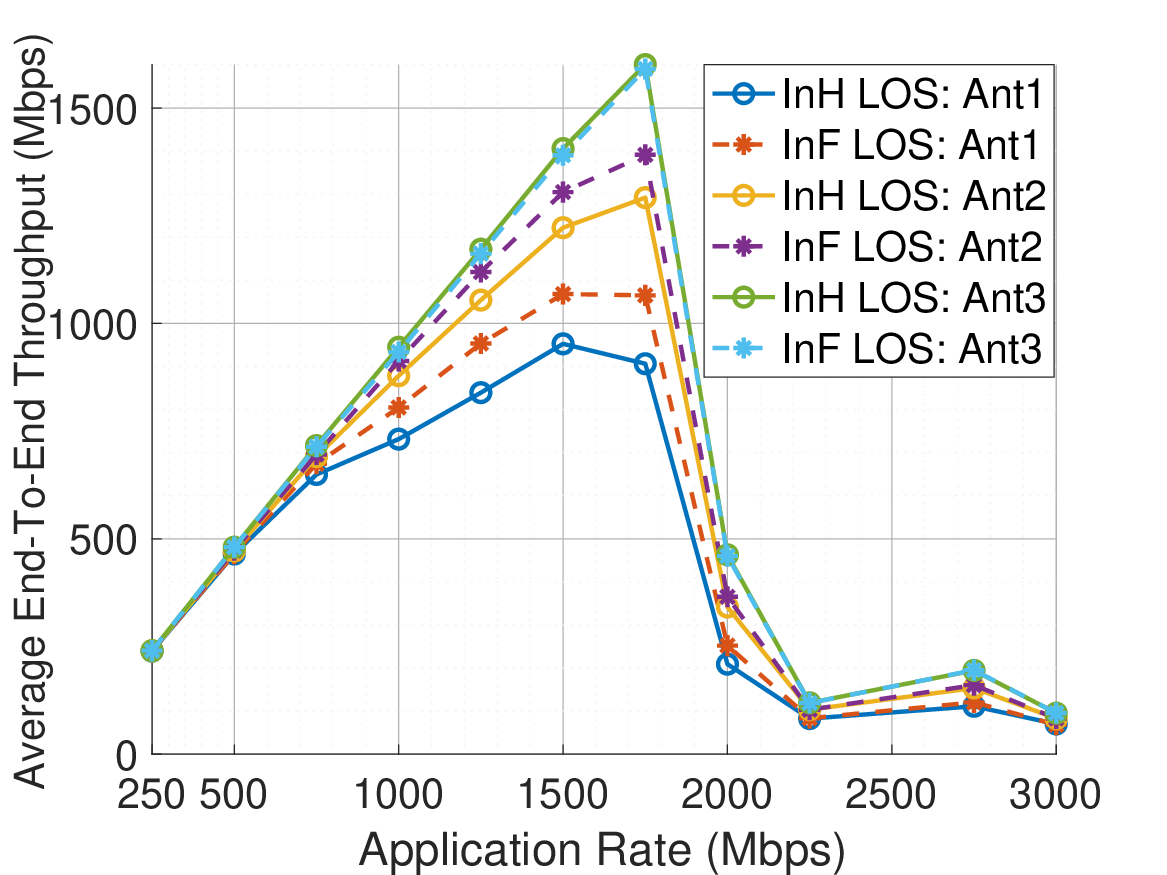}
        \caption{Average throughput vs application rate for 3 antenna configurations of gNB and UE in Indoor (InH and InF) LOS scenarios.}
        \label{fig:indoor_throughput_los}
    \end{subfigure}
    \hspace{0.2in}
    \begin{subfigure}[b]{0.2\textwidth}
        \centering
        \includegraphics[width=4.5cm,height=4.6cm,keepaspectratio]{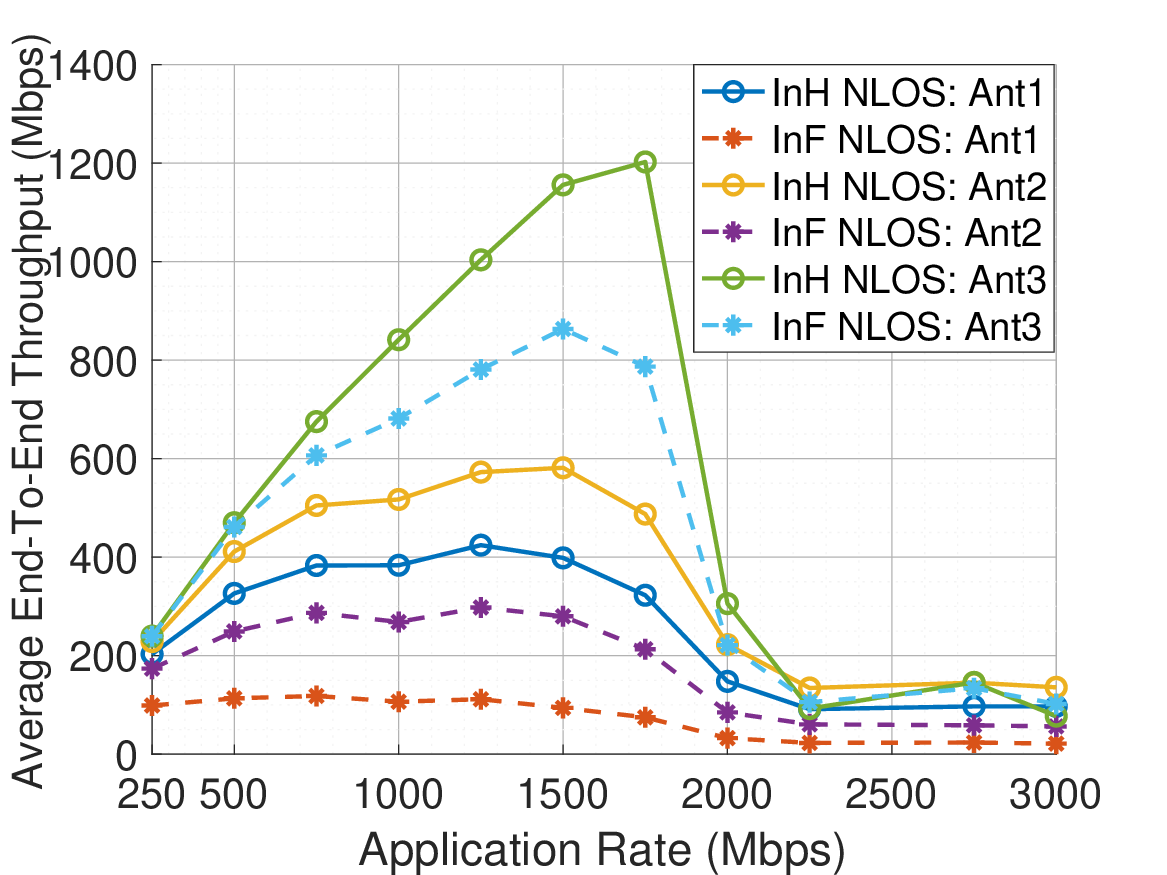}
        \caption{Average throughput vs Application rate 3 antenna configurations of gNB and UE in Indoor (InH and InF) NLOS scenarios.}
        \label{fig:indoor_throughput_nlos}
    \end{subfigure}
    \caption{Average downlink UE end-to-end throughput vs. application rate in Outdoor (UMi and RMa) and Indoor scenarios (InH and InF) for 100 m gNB-UE separation distance, 30 dBm transmit power, 140 GHz frequency, 1 GHz bandwidth and 2500 channel realizations.
    }
    \vspace{0.02in}
    \label{fig:throughput_1}
\end{figure*}

\begin{figure*}[h!]
    \centering
     \begin{subfigure}[b]{0.2\textwidth}
        \centering
\includegraphics[width=4.6cm,height=4.6cm,keepaspectratio]{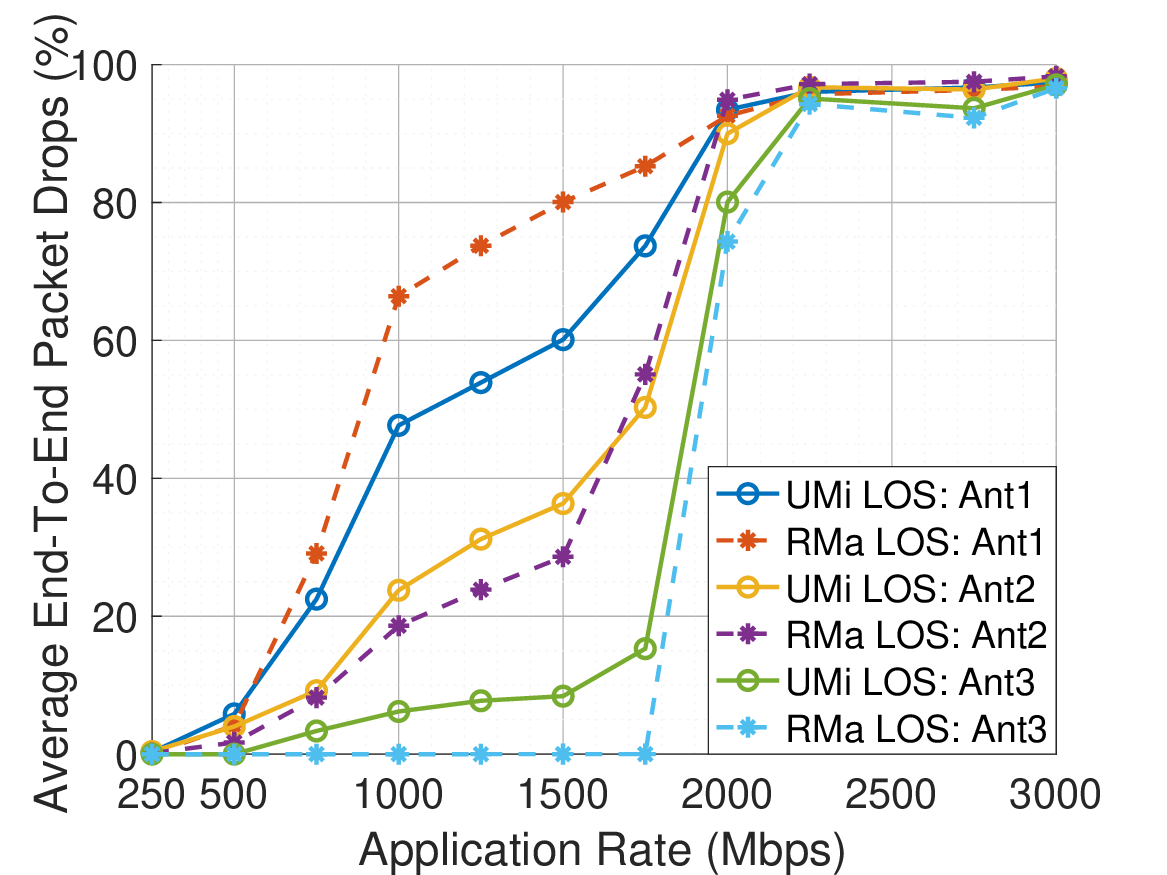}
        \caption{Average packet drop vs application rate for 3 antenna configurations of gNB and UE in Outdoor (UMi and RMa) LOS scenarios.}
        \label{fig:outdoor_packetdrop_los}
    \end{subfigure}
    \hspace{0.2in}
    \begin{subfigure}[b]{0.2\textwidth}
        \centering
        \includegraphics[width=4.6cm,height=4.6cm,keepaspectratio]{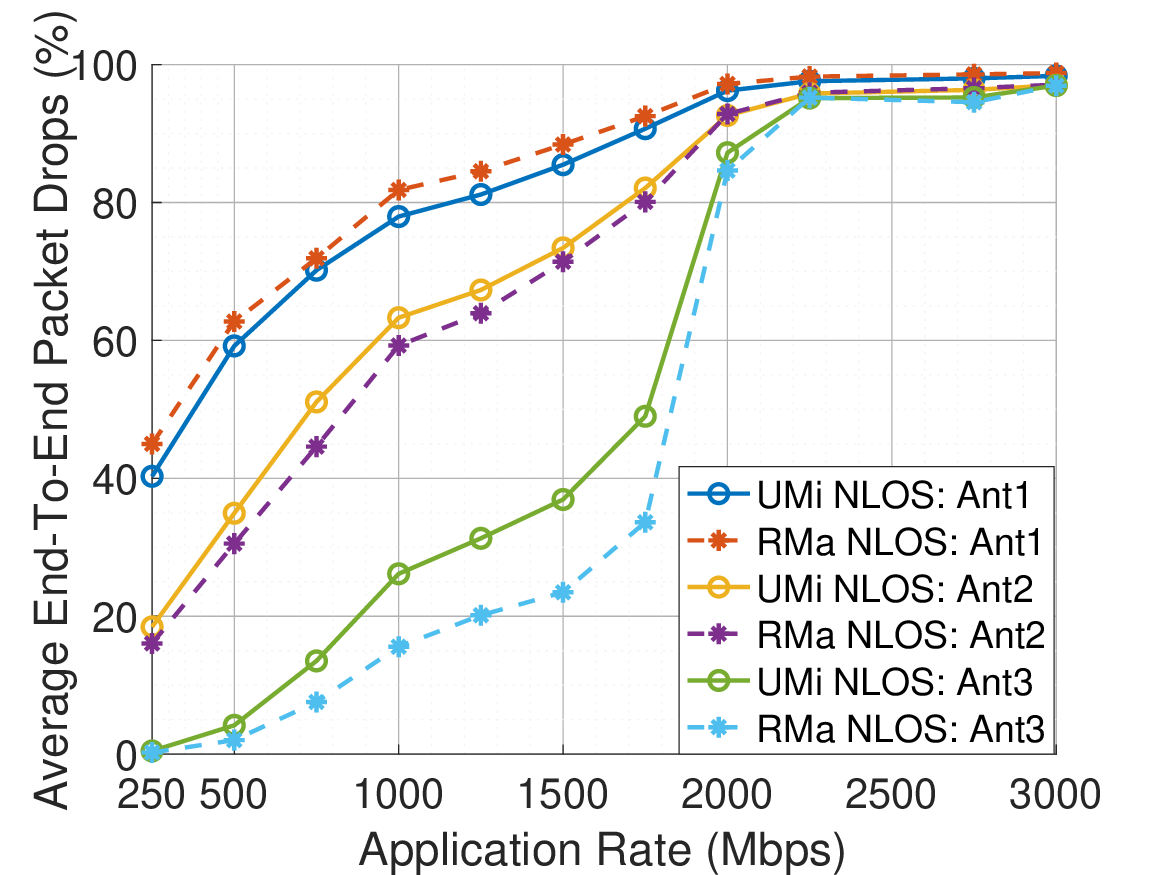}
        \caption{Average packet drop vs application rate for 3 antenna configurations of gNB and UE in Outdoor (UMi and RMa) NLOS scenarios.}
        \label{fig:outdoor_packetdrop_nlos}
    \end{subfigure}
    \hspace{0.2in}
    \begin{subfigure}[b]{0.2\textwidth}
        \centering
        \includegraphics[width=4.6cm,height=4.6cm,keepaspectratio]{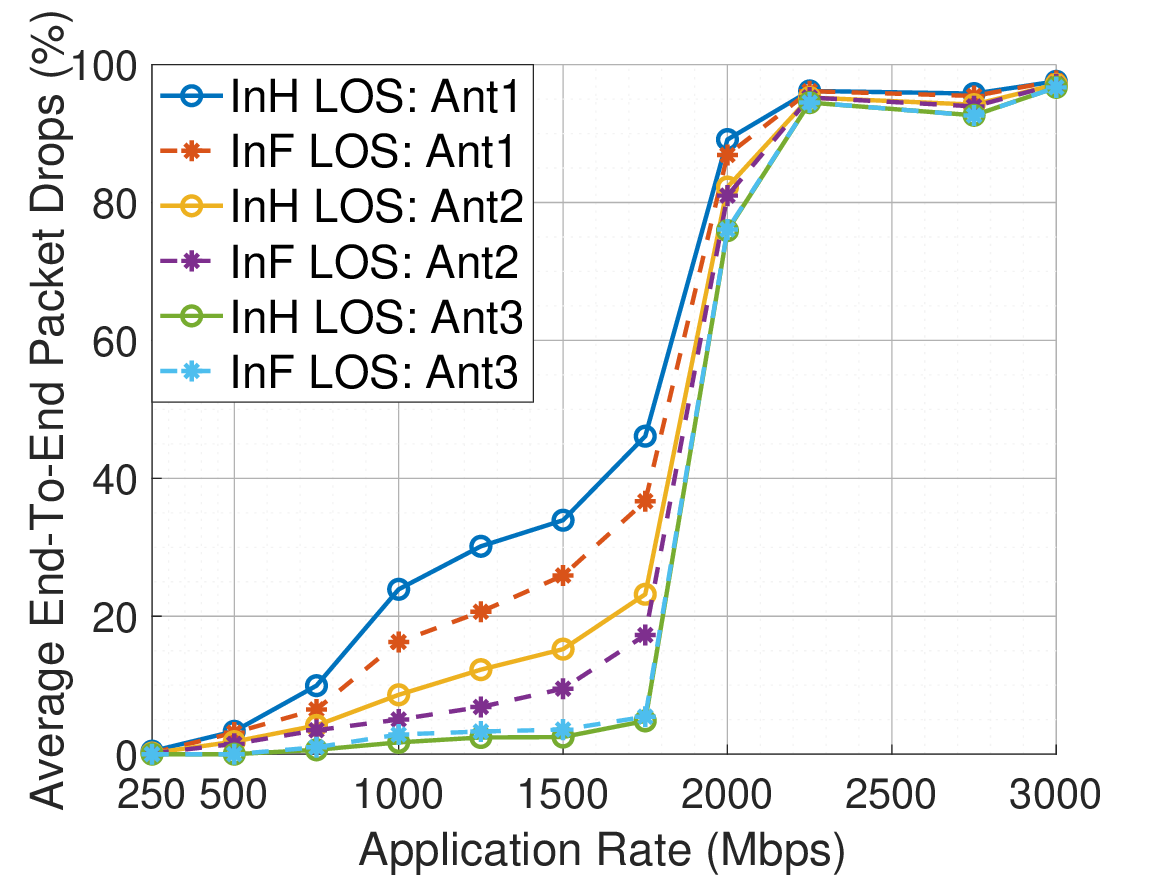}
        \caption{Average packet drop vs application rate for 3 antenna configurations of gNB and UE in Indoor (InH and InF) LOS scenarios.}
        \label{fig:indoor_packetdrop_los}
    \end{subfigure}
    \hspace{0.2in}
        \begin{subfigure}[b]{0.2\textwidth}
        \centering
        \includegraphics[width=4.5cm,height=4.6cm,keepaspectratio]{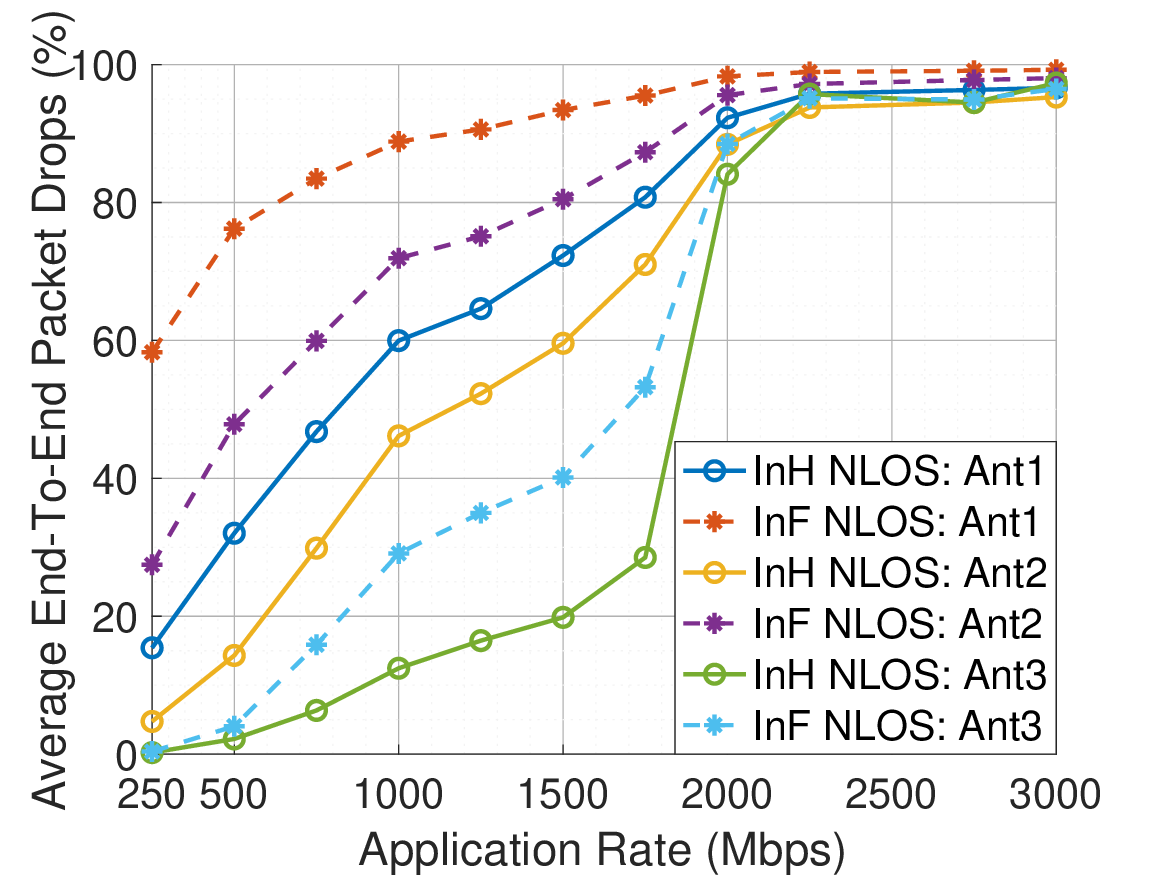}
        \caption{Average packet drop vs application rate for 3 antenna configurations of gNB and UE in Indoor (InH and InF) NLOS scenarios.}
        \label{fig:indoor_packetdrop_nlos}
    \end{subfigure}
    \caption{Average downlink UE end-to-end packet drops vs. application rate in Outdoor (UMi and RMa) and Indoor scenarios (InH and InF) for 100 m gNB-UE separation distance, 30 dBm transmit power, 140 GHz frequency, 1 GHz bandwidth and 2500 channel realizations.}
     \vspace{0.02in}
    \label{fig:packetdrops_1}
\end{figure*}
\begin{figure*}[htbp]
    \centering
    \begin{subfigure}[b]{0.2\textwidth}
        \centering
        \includegraphics[width=4.6cm,height=4.6cm,keepaspectratio]{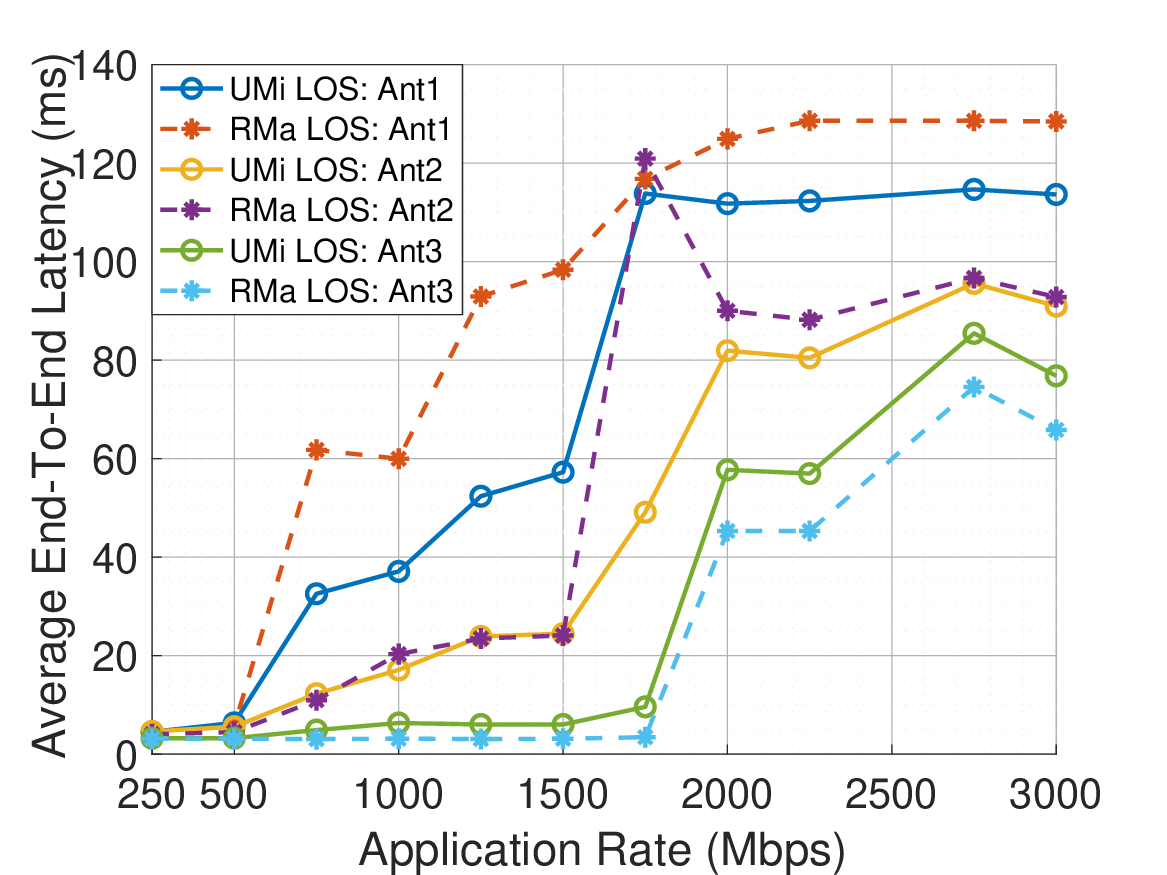}
        \caption{Average latency vs application rate for 3 antenna configurations of gNB and UE in Outdoor (UMi and RMa) LOS scenarios.}
        \label{fig:outdoor_latency_los}
    \end{subfigure}
    \hspace{0.2in}
        \begin{subfigure}[b]{0.2\textwidth}
        \centering
        \includegraphics[width=4.6cm,height=4.6cm,keepaspectratio]{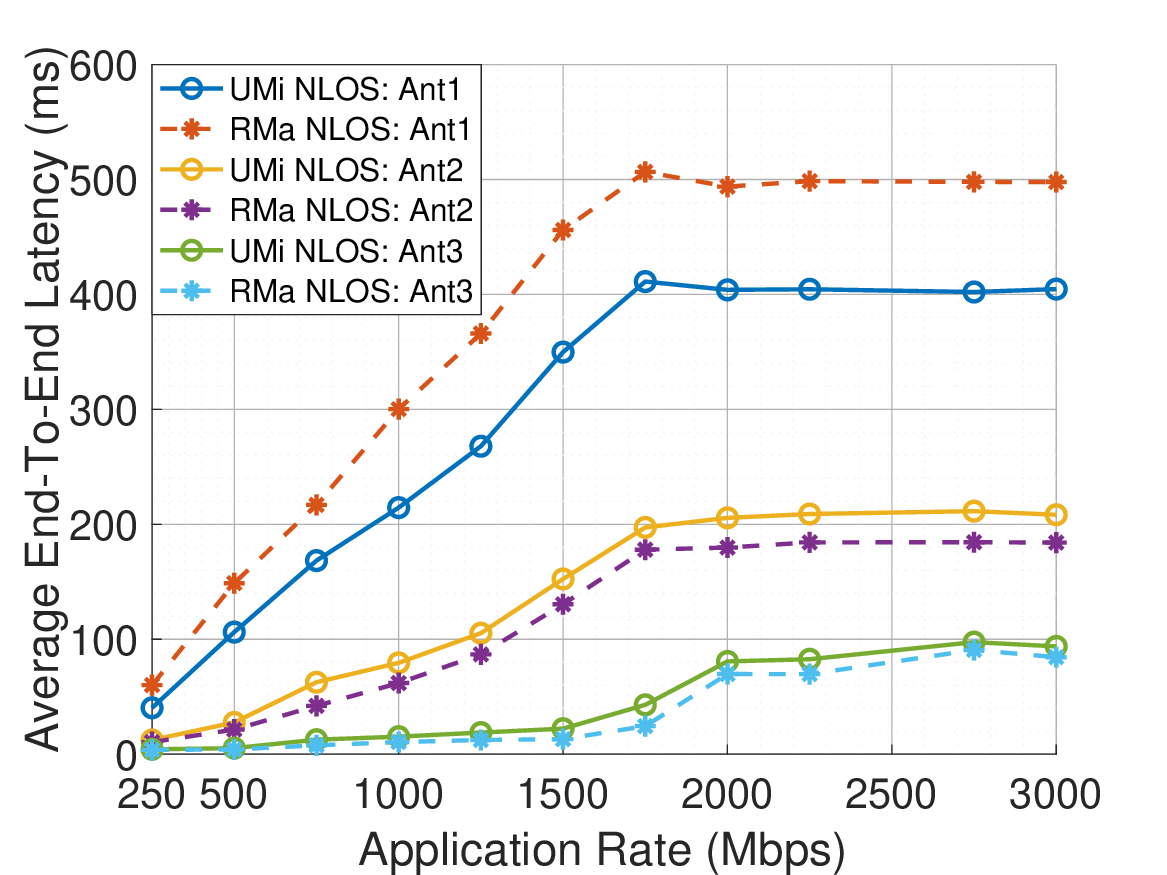}
        \caption{Average latency vs application rate for 3 antenna configurations of gNB and UE in Outdoor (UMi and RMa) NLOS scenarios.}
        \label{fig:outdoor_latency_nlos}
    \end{subfigure}
    \hspace{0.2in}
    \begin{subfigure}[b]{0.2\textwidth}
        \centering
        \includegraphics[width=4.6cm,height=4.6cm,keepaspectratio]{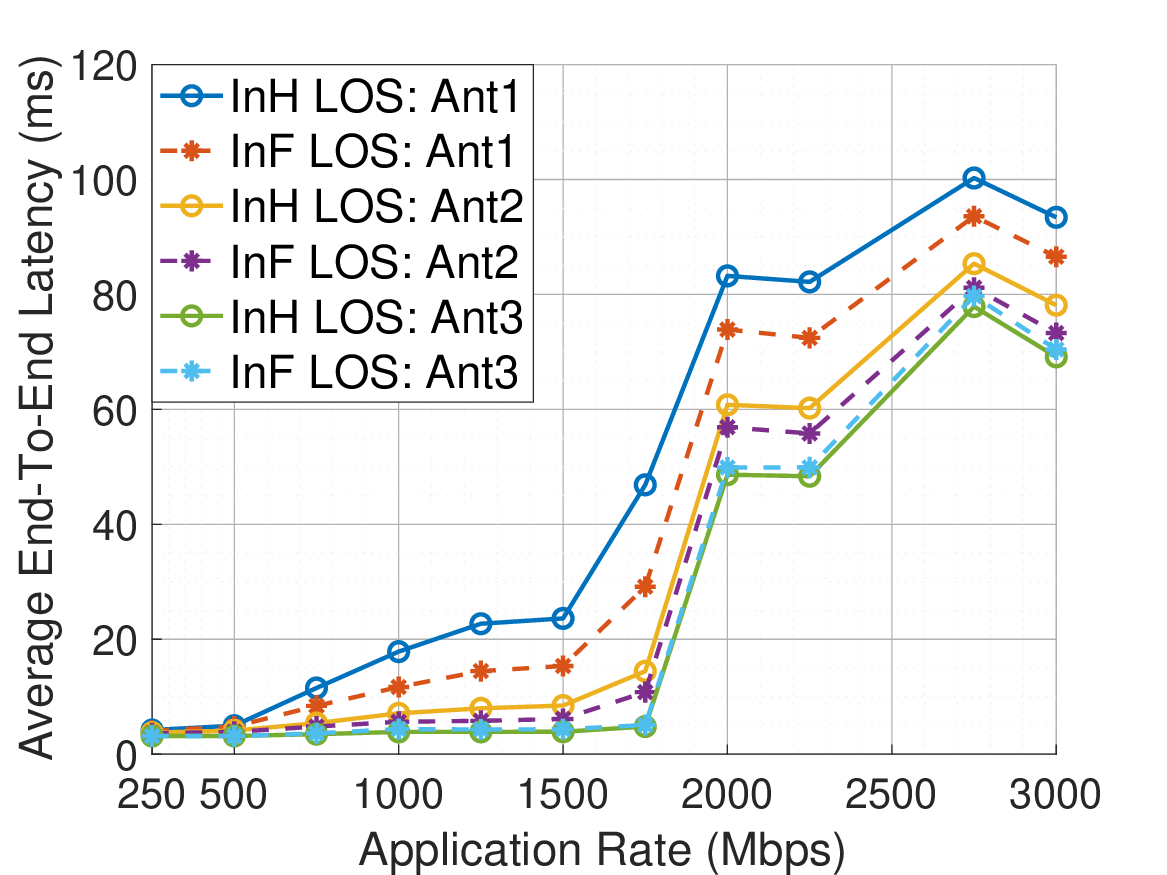}
        \caption{Average latency vs application rate for 3 antenna configurations of gNB and UE in Indoor (InH and InF) LOS scenarios.}
        \label{fig:indoor_latency_los}
    \end{subfigure}
        \hspace{0.2in}
    \begin{subfigure}[b]{0.2\textwidth}
        \centering
        \includegraphics[width=4.5cm,height=4.6cm,keepaspectratio]{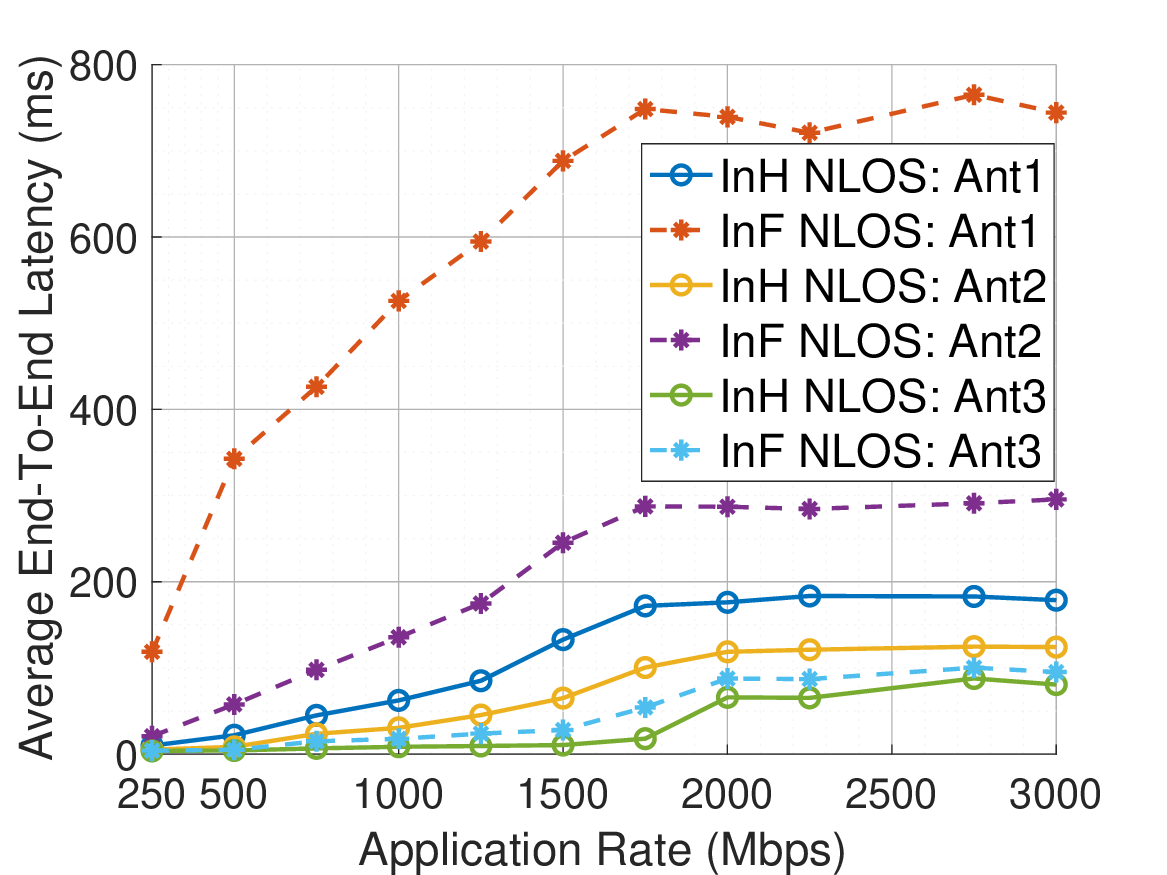}
        \caption{Average latency vs application rate for 3 antenna configurations of gNB and UE in Indoor (InH and InF) NLOS scenarios.}
        \label{fig:indoor_latency_nlos}
    \end{subfigure}
    \caption{Average downlink UE end-to-end latency vs. application rate in Outdoor (UMi and RMa) and Indoor scenarios (InH and InF) for 100 m gNB-UE separation distance, 30 dBm transmit power, 140 GHz frequency, 1 GHz bandwidth and 2500 realizations.}
    \vspace{-0.2in}
    \label{fig:latency}
\end{figure*}
We present the experimental results in two distinct categories: Outdoor and Indoor. The Outdoor category encompasses UMi and RMa scenarios, while the Indoor category includes InH and InF scenarios. We omit a detailed presentation of results for the UMa scenario as they closely align with those of UMi due to similar small-scale and large-scale channel parameters at 140 GHz \cite{poddar:2023:ns-3-implementation-of-sub}. In this paper, we use the term `SNR' to refer to the signal-to-noise ratio at the UE.
Analyzing Figure \ref{fig:outdoor_snr}, we observe that the SNR in the RMa scenario is consistently lower than in the UMi scenario, regardless of the channel condition and antenna configuration. The lower SNR in the RMa scenario is due to the higher path loss exponent compared to UMi under both channel conditions \cite{poddar:2023:ns-3-implementation-of-sub}. Additionally, increasing the number of antenna elements at the gNB and UE (Ant1, Ant2, Ant3) results in a rightward shift of the SNR curves for both the UMi and RMa scenarios, indicating that a higher number of antenna elements yields higher antenna gain, resulting in a strong received signal and consequently an increase in the SNR of the signal received at the UE. Similar trends are also evident for the Indoor (InH and InF) scenarios in Figure \ref{fig:indoor_snr} with increasing the number of antenna elements at the gNB and UE. However, the InH and InF scenarios exhibit approximately the same path loss exponent in LOS conditions and thus similar received power and SNR, which is reflected in the SNR curves of Figure \ref{fig:indoor_snr}. In contrast, the InF scenario experiences a higher path loss exponent in NLOS channel conditions \cite{poddar:2023:ns-3-implementation-of-sub} compared to the InH scenario, causing a lower SNR for the InF scenario compared to the InH scenario.
\par Figures \ref{fig:outdoor_throughput_los}, \ref{fig:outdoor_throughput_nlos}, \ref{fig:outdoor_packetdrop_los}, and \ref{fig:outdoor_packetdrop_nlos} illustrate the variation of average end-to-end throughput and average packet drops with respect to application rate for all three gNB-UE antenna configurations in LOS and NLOS channel conditions for the Outdoor scenarios (UMi and RMa). It is worth noting from Figures \ref{fig:outdoor_throughput_los} and \ref{fig:outdoor_throughput_nlos} that the throughput increases with the application rate until reaching a certain threshold for a given antenna configuration. The threshold for each antenna configuration is determined by the maximum physical layer throughput that can be sustained by the link \cite{poddar:2023:full-stack}. The maximum physical layer throughput can be calculated using the transport block (TB) size and slot duration where the TB size is in bits. For instance, consider a sample run from the UMi LOS scenario (Figure \ref{fig:outdoor_throughput_los}) using antenna configuration ``Ant3'' having the maximum possible Modulation and Coding Scheme (MCS) of 28 and TB size 56200 bytes. For this case, the maximum possible physical layer throughput calculated from \cite{poddar:2023:full-stack} will be $\sim$1800 Mbps. Thus, when the source application rate (values in the x-axis of Figure \ref{fig:outdoor_throughput_los}) exceeds the maximum physical layer throughput, a drastic decline in throughput is observed which can be attributed to buffer overflows, as explained in \cite{poddar:2023:full-stack}. Additionally, increasing source application rates beyond the maximum sustainable physical layer throughput leads to excessive packet drops as seen in Figures \ref{fig:outdoor_packetdrop_los} and \ref{fig:outdoor_packetdrop_nlos}. Moreover, when conducting 2500 realizations using a particular antenna configuration, source application rate, scenario, and channel condition, it's important to note that not all realizations will have strong SNRs. This is because each realization generates an independent channel and thus a different SNR. In realizations where the SNR is low, the physical layer throughput is low, leading to an increase in packet loss caused by buffer overflows \cite{poddar:2023:full-stack}. Note that in practical end-to-end communication scenarios, packet drop can stem from various sources, not just a buffer overflow. However, in this particular simulation scenario, the root cause of packet drop is due to buffer overflows. By taking the average of all realizations, we calculate the average end-to-end throughput and packet drops. Interestingly, even before reaching the threshold value, we observe a rise in packet loss and a decline in throughput. The reason behind this is that some of the realizations cannot sustain such high source application rates, resulting in degradation. For instance, from Figure \ref{fig:outdoor_throughput_los} in Ant 3 configuration we observe that RMa LOS outperforms UMi LOS, this is because of some of the realizations for UMi LOS in Ant 3 that have extremely bad SNR and incur higher packet drops, and thus the averaging overall realizations leads to an overall decrease in throughput. Moreover, if the application source rate exceeds the maximum sustained physical layer throughput, the reduction in throughput becomes consistent. For instance, as demonstrated in Figure \ref{fig:outdoor_throughput_los}, it becomes evident that the ``Ant3'' configuration cannot sustain a source application rate beyond 1800 Mbps. Consequently, any attempt to increase the source application rate beyond 1800 Mbps consistently results in a significant drop in throughput. These findings highlight a notable observation: in certain scenarios characterized by low SNR in specific realizations, the physical layer throughput exhibits a decline, even when the source application rates are set at relatively lower levels. This observation underscores the fact that in such scenarios, throughput can decrease, irrespective of the source application rate being set at modest levels. Similar conclusions can be drawn from Figures \ref{fig:indoor_throughput_los}, \ref{fig:indoor_throughput_nlos}, \ref{fig:indoor_packetdrop_los}, and \ref{fig:indoor_packetdrop_nlos}, which illustrate average end-to-end throughput and packet drops in the Indoor (InH and InF) scenarios under LOS and NLOS channel conditions, respectively.
\par Figure \ref{fig:latency} depicts the latency for the Outdoor (UMi and RMa) and Indoor (InH and InF) scenarios under LOS and NLOS channel conditions. We can observe from Figure \ref{fig:latency} that latency is related to packet drops. As the packet drop increases, the latency also increases. This phenomenon can be attributed to the activation of HARQ triggered by packet drops. HARQ aims to reduce end-to-end packet drops and enhance end-to-end throughput at the expense of increased latency \cite{poddar:2023:full-stack}. 

%% file: Sections/Tradeoff.tex
\begin{table*}[htbp]
\centering
  \caption{\centering The observed throughput (Mbps) and packet drop (\%) for a latency target of less than 5 ms and 10 ms using all three antenna configurations (Ant1, Ant2 and Ant3) for outdoor (UMi and RMa) scenarios in LOS channel condition.}
    \begin{tabular}{rrrrrrrrrrrrr} \\
    \hline
    \multicolumn{1}{|c|}{\multirow{3}[6]{*}{\textbf{Latency Target (ms)}}} & \multicolumn{6}{c|}{\textbf{Outdoor LOS Throughput (Mbps)}} & \multicolumn{6}{c|}{\textbf{Outdoor LOS Packet Drop (\%)}} \\
\cline{2-13}    \multicolumn{1}{|c|}{} & \multicolumn{2}{c|}{Ant1} & \multicolumn{2}{c|}{Ant2} & \multicolumn{2}{c|}{Ant3} & \multicolumn{2}{c|}{Ant1} & \multicolumn{2}{c|}{Ant2} & \multicolumn{2}{c|}{Ant3} \\
\cline{2-13}    \multicolumn{1}{|c|}{} & \multicolumn{1}{c|}{UMi} & \multicolumn{1}{c|}{RMa} & \multicolumn{1}{c|}{UMi} & \multicolumn{1}{c|}{RMa} & \multicolumn{1}{c|}{UMi} & \multicolumn{1}{c|}{RMa} & \multicolumn{1}{c|}{UMi} & \multicolumn{1}{c|}{RMa} & \multicolumn{1}{c|}{UMi} & \multicolumn{1}{c|}{RMa} & \multicolumn{1}{c|}{UMi} & \multicolumn{1}{c|}{RMa} \\
    \hline
    \multicolumn{1}{|c|}{< 5 ms} & \multicolumn{1}{c|}{500} & \multicolumn{1}{c|}{500} & \multicolumn{1}{c|}{500} & \multicolumn{1}{c|}{500} & \multicolumn{1}{c|}{1500} & \multicolumn{1}{c|}{1500} & \multicolumn{1}{c|}{5} & \multicolumn{1}{c|}{5} & \multicolumn{1}{c|}{5} & \multicolumn{1}{c|}{5} & \multicolumn{1}{c|}{0} & \multicolumn{1}{c|}{0} \\
    \hline
    \multicolumn{1}{|c|}{< 10 ms} & \multicolumn{1}{c|}{500} & \multicolumn{1}{c|}{500} & \multicolumn{1}{c|}{750} & \multicolumn{1}{c|}{750} & \multicolumn{1}{c|}{1750} & \multicolumn{1}{c|}{1750} & \multicolumn{1}{c|}{5} & \multicolumn{1}{c|}{5} & \multicolumn{1}{c|}{10} & \multicolumn{1}{c|}{10} & \multicolumn{1}{c|}{15} & \multicolumn{1}{c|}{0} \\
    \hline  \\
    \end{tabular}%
  \label{tab:outdoorlos}%
\end{table*}%
\begin{table*}[htbp]
\centering
  \caption{\centering The observed throughput (Mbps) and packet drop (\%) for a latency target of less than 5 ms and 10 ms using all three antenna configurations (Ant1, Ant2 and Ant3) for outdoor (UMi and RMa) scenarios in NLOS channel condition. }
    \begin{tabular}{|c|c|c|c|c|c|c|c|c|c|c|c|c|}
    \hline
    \multirow{3}[6]{*}{\textbf{Latency Target (ms)}} & \multicolumn{6}{c|}{\textbf{Outdoor NLOS Throughput (Mbps)}} & \multicolumn{6}{c|}{\textbf{Outdoor NLOS Packet Drop (\%)}} \\
\cline{2-13}          & \multicolumn{2}{c|}{Ant1} & \multicolumn{2}{c|}{Ant2} & \multicolumn{2}{c|}{Ant3} & \multicolumn{2}{c|}{Ant1} & \multicolumn{2}{c|}{Ant2} & \multicolumn{2}{c|}{Ant3} \\
\cline{2-13}          & UMi   & RMa   & UMi   & RMa   & UMi   & RMa   & UMi   & RMa   & UMi   & RMa   & UMi   & RMa \\
    \hline
    < 5 ms & < 250 & < 250 & < 250 & < 250 & 1000  & 1000  & < 40  & < 42  & < 18  & < 16  & 25    & 15 \\
    \hline
    < 10 ms & < 250 & < 250 & 500   & 500   & 1500  & 1500  & < 40  & < 42  & 35    & 30    & 38    & 22 \\
    \hline
    \end{tabular}%
  \label{tab:outdoornlos}%
\end{table*}%
\begin{table*}[htbp]
\centering
  \caption{\centering The observed throughput (Mbps) and packet drop (\%) for a latency target of less than 5 ms and 10 ms using all three antenna configurations (Ant1, Ant2 and Ant3) for indoor (InH and InF) scenarios in LOS channel condition.}
    \begin{tabular}{|c|c|c|c|c|c|c|c|c|c|c|c|c|}
    \hline
    \multirow{3}[6]{*}{\textbf{Latency Target (ms)}} & \multicolumn{6}{c|}{\textbf{Indoor LOS Throughput (Mbps)}} & \multicolumn{6}{c|}{\textbf{Indoor LOS Packet Drop (\%)}} \\
\cline{2-13}          & \multicolumn{2}{c|}{Ant1} & \multicolumn{2}{c|}{Ant2} & \multicolumn{2}{c|}{Ant3} & \multicolumn{2}{c|}{Ant1} & \multicolumn{2}{c|}{Ant2} & \multicolumn{2}{c|}{Ant3} \\
\cline{2-13}          & InH   & InF   & InH   & InF   & InH   & InF   & InH   & InF   & InH   & InF   & InH   & InF \\
    \hline
    < 5 ms & 550   & 550   & 1250  & 1500  & 1800  & 1800  & 5     & 5     & 11    & 8     & 6     & 6 \\
    \hline
    < 10 ms & 750   & 1200  & 1750  & 1800  & 1850  & 1850  & 10    & 8     & 22    & 18    & 20    & 20 \\
    \hline
    \end{tabular}%
  \label{tab:indoorlos}%
\end{table*}%
\begin{table*}[htbp]
  \centering
  \caption{\centering The observed throughput (Mbps) and packet drop (\%) for a latency target of less than 5 ms and 10 ms using all three antenna configurations (Ant1, Ant2 and Ant3) for indoor (InH and InF) scenarios in NLOS channel condition.}
    \begin{tabular}{|c|c|c|c|c|c|c|c|c|c|c|c|c|}
    \hline
    \multirow{3}[6]{*}{\textbf{Latency Target (ms)}} & \multicolumn{6}{c|}{\textbf{Indoor NLOS Throughput (Mbps)}} & \multicolumn{6}{c|}{\textbf{Indoor NLOS Packet Drop (\%)}} \\
\cline{2-13}          & \multicolumn{2}{c|}{Ant1} & \multicolumn{2}{c|}{Ant2} & \multicolumn{2}{c|}{Ant3} & \multicolumn{2}{c|}{Ant1} & \multicolumn{2}{c|}{Ant2} & \multicolumn{2}{c|}{Ant3} \\
\cline{2-13}          & InH   & InF   & InH   & InF   & InH   & InF   & InH   & InF   & InH   & InF   & InH   & InF \\
    \hline
    < 5 ms & < 250 & < 250 & 250   & 250   & 500   & 500   & < 15  & < 60  & 5     & 28    & 2     & 3 \\
    \hline
    < 10 ms & 250   & 250   & 500   & 300   & 1750  & 1250  & 15    & 60    & 12    & 30    & 15    & 15 \\
    \hline
    \end{tabular}%
  \label{tab:indoornlos}%
\end{table*}%
This subsection aims to provide an understanding of how a chosen KPI influences other performance metrics. In Tables \ref{tab:outdoorlos}, \ref{tab:outdoornlos}, \ref{tab:indoorlos}, and \ref{tab:indoornlos} by specifying a desired latency target of less than 5 and 10 milliseconds (ms)  for outdoor and indoor scenarios in LOS and NLOS channel conditions, we showcase the corresponding levels of throughput and packet drop that are observed. A report \cite{qualcommVideo} published by Qualcomm highlighted that certain applications in 6G which include extended reality (XR), autonomous vehicles, crowded event sharing, remote control, and immersive content require data rates exceeding 1 Gbps and latency below 15 ms. Our analysis from Tables \ref{tab:outdoorlos} and \ref{tab:outdoornlos} demonstrates that, in outdoor LOS and NLOS scenarios, the sub-THz band can enable applications demanding over 1 Gbps and latency under 15 ms. This is achievable using the existing mmWave protocol stack but with a large antenna array (in this paper denoted as ``Ant3''). Conversely, for indoor LOS scenario, the same benchmarks can be met using ``Ant2'' as shown in Table \ref{tab:indoorlos}. However, in indoor NLOS scenario, the stipulated throughput and latency criteria are achieved using ``Ant3'' as seen in Table \ref{tab:indoornlos}. Tables \ref{tab:outdoornlos} and \ref{tab:indoornlos} do not provide exact numerical values for some of the performance metrics, as the corresponding Figures \ref{fig:throughput_1}, \ref{fig:packetdrops_1}, and \ref{fig:latency} do not fully reveal the granular values due to the wide range of application source rates (250 Mbps to 3000 Mbps) used in the current simulation. To capture exact values, it is recommended that readers simulate with narrower application source rate ranges, such as 10--250 Mbps. This analysis serves to highlight the trade-offs and dependencies that exist among various performance metrics. Note that even though we use latency as a KPI in this case, one could have chosen other metrics like throughput or packet drop. For instance for a desired throughput of 500 Mbps, one could estimate what is the observed latency and packet drop. By gaining insights into how selecting a KPI can provide information about other metrics, engineers can make well-informed decisions and establish realistic expectations regarding system performance in real-world scenarios.

%% file: Sections/Impact.tex
Performing extensive Monte Carlo wireless simulations for system-level or network-level analyses for any frequency, scenario, etc can pose significant computational and time constraints. Typically, statistical significance in Monte Carlo simulations is achieved by $10^4$--$10^6$ realizations \cite{tranter:2004:principles}. The end-to-end performance metrics are heavily dependent on the number of realizations to obtain statistically accurate results. By harnessing the capabilities of high-performance clusters (HPCs) and employing parallelization techniques tools such as ``sem'' \cite{magrin:2019:simulation} in ns-3 it becomes possible to complete $10^4$--$10^6$ realizations in comparatively much less time compared to standard workstations. However, determining the exact number of realizations required to obtain statistically significant results remains challenging. As a result, researchers commonly employ a range of realizations, spanning from a few hundred for expedited results to a vast number in the order of $10^4$ or more. Researchers must balance computational resources, time constraints, and the desired level of statistical confidence when designing their simulation experiments. Thus, in this work, we present the statistics of the number of realizations versus confidence intervals and strongly encourage researchers worldwide, utilizing ns-3 or any other simulation tools involving statistical models, to provide similar metrics in their work. Such metrics enable readers to assess the reported results' accuracy and reliability and enhance the finding's overall interpretability.

\begin{figure}[t!]
    \centering
    \includegraphics[width=9cm,height=10cm,keepaspectratio]{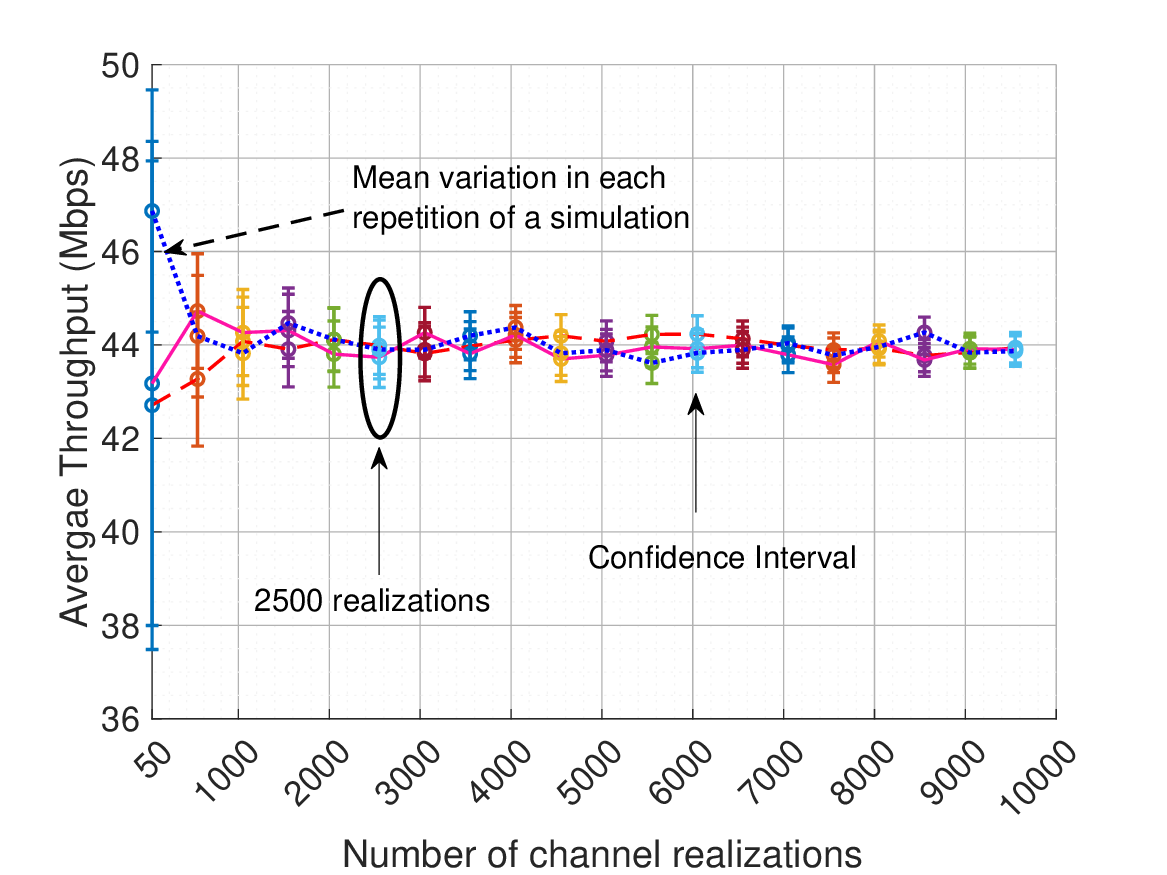}
    \caption{Analyzing the impact of the number of channel realizations on mean value and confidence intervals for throughput in UMi scenario, NLOS channel condition, with Ant3 configuration, and a fixed application source rate of 50 Mbps (3 simulations each with a duration of 9 seconds are conducted).}
    \vspace{-0.2in}
    \label{fig:confidence_intervals}
\end{figure}

Figure \ref{fig:confidence_intervals} illustrates the relationship between the number of realizations and the corresponding confidence intervals. Observing the graph, it is evident that with a lower number of realizations, the confidence interval is wider, indicating a higher degree of uncertainty. Moreover, the mean values exhibit significant variation within the confidence intervals, indicating instability in the results. Additionally, for a given number of realizations, the confidence interval shifts noticeably, which is undesirable for reliable analysis. As the number of realizations increases, the confidence interval becomes narrower, indicating increased precision and reduced uncertainty. The variation in mean values within the confidence intervals also decreases. Notably, in this particular experiment with 10,000 realizations, it can be observed that after approximately 2,000 realizations, the confidence interval stabilizes. The narrower confidence interval and reduced mean variation provide greater confidence in the obtained results. Based on these observations and time constraints, we have chosen to perform 2,500 realizations for each scenario, channel condition, source application rate, and antenna configuration ensuring statistically significant results with a reliable level of precision. 

%% file: Sections/Conclusion.tex
By conducting extensive single-user simulations using the ns-3 mmWave module, employing the NYUSIM channel model at 140 GHz, we obtain insights into end-to-end downlink SNR, average throughput, packet drops, and latency across various gNB and UE antenna configurations in different 3GPP-listed LOS and NLOS scenarios. \textit{Our simulations indicate that sub-THz bands can support applications demanding data rates above 1 Gbps and latency below 15 ms \cite{qualcommVideo}, leveraging the current mmWave protocol stack but with large antenna arrays}. Moreover, our findings reveal that increasing antenna elements at the gNB and UE enhances the 140 GHz system's performance by improving SNR. This SNR impacts MCS, influencing the transport block (TB) size and setting the receiver's maximum physical layer throughput for a given antenna configuration. Overshooting this throughput can lead to RLC buffer overflow, causing significant packet loss, latency spikes, and frequent retransmissions. Our findings highlight that the 8x8–4x4 antenna setup (Ant1) at 140 GHz is less efficient than its 28 GHz counterpart \cite{poddar:2023:full-stack}, emphasizing the necessity to adjust antenna elements to counteract heightened path loss after the first meter of propagation at 140 GHz. While our current study emphasizes single-user scenarios to obtain maximal performance insights, future research should delve into multi-user environments to capture real-world performance nuances, taking into account factors such as co/adjacent channel interference and blockage effects.

%% file: main.bbl
\begin{thebibliography}{10}
\providecommand{\url}[1]{#1}
\csname url@samestyle\endcsname
\providecommand{\newblock}{\relax}
\providecommand{\bibinfo}[2]{#2}
\providecommand{\BIBentrySTDinterwordspacing}{\spaceskip=0pt\relax}
\providecommand{\BIBentryALTinterwordstretchfactor}{4}
\providecommand{\BIBentryALTinterwordspacing}{\spaceskip=\fontdimen2\font plus
\BIBentryALTinterwordstretchfactor\fontdimen3\font minus \fontdimen4\font\relax}
\providecommand{\BIBforeignlanguage}[2]{{%
\expandafter\ifx\csname l@#1\endcsname\relax
\typeout{** WARNING: IEEEtran.bst: No hyphenation pattern has been}%
\typeout{** loaded for the language `#1'. Using the pattern for}%
\typeout{** the default language instead.}%
\else
\language=\csname l@#1\endcsname
\fi
#2}}
\providecommand{\BIBdecl}{\relax}
\BIBdecl

\bibitem{viswanathan:2020:communications-in-the}
H.~Viswanathan and P.~E. Mogensen, ``Communications in the {6G} era,'' \emph{IEEE Access}, vol.~8, pp. 57\,063--57\,074, 2020.

\bibitem{aazhang:2019:key-drivers}
\BIBentryALTinterwordspacing
B.~Aazhang \emph{et~al.}, ``{Key drivers and research challenges for {6G} ubiquitous wireless intelligence (white paper)},'' Univ. of Oulu, Finland, Tech. Rep. 978-952-62-2354-4, Sep 2019. [Online]. Available: \url{http://jultika.oulu.fi/files/isbn9789526223544.pdf}
\BIBentrySTDinterwordspacing

\bibitem{dang:2020:what-should}
S.~Dang \emph{et~al.}, ``What should {6G} be?'' \emph{Nature Electronics}, vol.~3, no.~1, pp. 20--29, 2020.

\bibitem{bazzi:2023:on-integrated-sensing}
A.~Bazzi and M.~Chafii, ``{On Integrated Sensing and Communication Waveforms With Tunable PAPR},'' \emph{IEEE Transactions on Wireless Communications}, vol.~22, no.~11, pp. 7345--7360, 2023.

\bibitem{rappaport:2019:wireless-communications-and-applications}
T.~S. Rappaport \emph{et~al.}, ``{Wireless Communications and Applications Above 100 GHz: Opportunities and Challenges for 6G and Beyond},'' \emph{IEEE Access}, vol.~7, pp. 78\,729--78\,757, 2019.

\bibitem{chaffi:2023:twelve-scientific-challenges}
M.~Chafii \emph{et~al.}, ``{Twelve Scientific Challenges for 6G: Rethinking the Foundations of Communications Theory},'' \emph{IEEE Communications Surveys \& Tutorials}, vol.~25, no.~2, pp. 868--904, 2023.

\bibitem{rappaport:2013:millimeter-wave-mobile}
T.~S. Rappaport \emph{et~al.}, ``{Millimeter wave mobile communications for {5G} cellular: It will work!}'' \emph{IEEE Access}, vol.~1, pp. 335--349, 2013.

\bibitem{naga:2014:network-densification}
N.~Bhushan \emph{et~al.}, ``{N}etwork densification: the dominant theme for wireless evolution into {5G},'' \emph{IEEE Communications Magazine}, vol.~52, no.~2, pp. 82--89, 2014.

\bibitem{next-g-alliance:2022:next-g-alliance-report}
\BIBentryALTinterwordspacing
{Next G Alliance}, ``{Next G Alliance Report: Roadmap to 6G},'' Washington, DC, USA, Tech. Rep., Feb 2022. [Online]. Available: \url{https://www.nextgalliance.org/wp-content/uploads/2022/02/NextGA-Roadmap.pdf}
\BIBentrySTDinterwordspacing

\bibitem{fcc}
\BIBentryALTinterwordspacing
{Federal Communications Commission}, ``{Spectrum Horizons--First Report and Order--ET Docket No. 18--21},'' Washington, DC, USA, March 2019. [Online]. Available: \url{https://docs.fcc.gov/public/attachments/FCC-19-19A1.pdf}
\BIBentrySTDinterwordspacing

\bibitem{singh:2019:beyond-5g-THz-spectrum}
R.~Singh and D.~Sicker, ``Beyond {5G}: {THz} {Spectrum} {Futures} and {Implications} for {Wireless} {Communication},'' in \emph{30th {European} {Conference} of the {International} {Telecommunications} {Society} ({ITS})}, 2019, pp. 1--30.

\bibitem{chen:2023:5g-advacned-towards-6g}
W.~Chen \emph{et~al.}, ``{5G-Advanced Toward 6G: Past, Present, and Future},'' \emph{IEEE Journal on Selected Areas in Communications}, vol.~41, no.~6, pp. 1592--1619, 2023.

\bibitem{mezzavilla:2018:end-to-end-simulation-of-5g-mmwave}
M.~Mezzavilla \emph{et~al.}, ``End-to-end simulation of {5G} mmwave networks,'' \emph{IEEE Communication Surveys \& Tutorials}, vol.~20, no.~3, pp. 2237--2263, 2018.

\bibitem{3gpp:2022:3gpp-ts-38.101-1}
\BIBentryALTinterwordspacing
{3GPP}, ``{NR}; {U}ser {E}quipment {(UE)} radio transmission and reception; {P}art 1: {R}ange 1 {S}tandalone,'' Tech. Rep. TS--38.101--1--Ver.15.20.0, January 2023. [Online]. Available: \url{https://www.3gpp.org/ftp/Specs/archive/38_series/38.101-1/38101-1-fk0.zip}
\BIBentrySTDinterwordspacing

\bibitem{3gpp:2022:3gpp-ts-38.101-2}
3GPP, ``{NR}; {U}ser {E}quipment {(UE)} radio transmission and reception; {P}art 2: {R}ange 2,'' Tech. Rep. TS--38.101--2, 2023.

\bibitem{poddar:2023:ns-3-implementation-of-sub}
H.~Poddar \emph{et~al.}, ``{ns-3 Implementation of Sub-Terahertz and Millimeter Wave Drop-Based NYU Channel Model (NYUSIM)},'' in \emph{Workshop on ns-3}, New York, NY, USA, 2023, pp. 19--27.

\bibitem{poddar:2023:full-stack}
{H. Poddar} \emph{et~al.}, ``{Full-Stack End-to-End mmWave Simulations Using 3GPP and NYUSIM Channel Model in ns-3},'' in \emph{IEEE Int. Conf. Commun. (ICC)}, 2023, pp. 1048--1053.

\bibitem{poddar:2023:nyusim-in-ns3}
\BIBentryALTinterwordspacing
H.~Poddar \emph{et~al.}, ``{NYUSIM}: {mmWave} and sub-{THz} channel simulator in ns-3,'' NYU WIRELESS, 2023. [Online]. Available: \url{https://apps.nsnam.org/app/nyusim/}
\BIBentrySTDinterwordspacing

\bibitem{sun:2016:nyusim:-mmwave-and-sub-thz}
\BIBentryALTinterwordspacing
{H. Poddar} \emph{et~al.}, ``{NYUSIM}: {mmWave} and sub-{THz} channel simulator in {MATLAB},'' NYU WIRELESS, 2023. [Online]. Available: \url{https://wireless.engineering.nyu.edu/nyusim/}
\BIBentrySTDinterwordspacing

\bibitem{sun:2017:a-novel-millimeter-wave-channel}
S.~Sun \emph{et~al.}, ``{A Novel Millimeter-Wave Channel Simulator and Applications for {5G} Wireless Communications},'' in \emph{IEEE International Conference on Communications (ICC)}, May 2017, pp. 1--7.

\bibitem{ju:2019:a-millimeter-wave-channel-simulator}
S.~Ju \emph{et~al.}, ``{A Millimeter-wave Channel Simulator {NYUSIM} with Spatial Consistency and Human Blockage},'' in \emph{IEEE Global Communications Conference (GLOBECOM)}, 2019, pp. 1--6.

\bibitem{poddar:2023:a-tutorial-on-nyusim}
H.~Poddar \emph{et~al.}, ``{A Tutorial on {NYUSIM}: Sub-Terahertz and Millimeter-Wave Channel Simulator for 5G, 6G and Beyond},'' \emph{IEEE Communications Surveys \& Tutorials}, pp. 1--1, 2023.

\bibitem{rappaport:2013:broadband-millimeter-wave-propagation}
T.~S. Rappaport \emph{et~al.}, ``{Broadband Millimeter-Wave Propagation Measurements and Models Using Adaptive-Beam Antennas for Outdoor Urban Cellular Communications},'' \emph{IEEE Transactions on Antennas and Propagation}, vol.~61, no.~4, pp. 1850--1859, Apr. 2013.

\bibitem{rappaport:2015:wideband-millimeter-wave-propagation}
{T. S. Rappaport} \emph{et~al.}, ``{Wideband Millimeter-wave Propagation Measurements and Channel Models for Future Wireless Communication System Design},'' \emph{IEEE Transactions on Communications}, vol.~63, no.~9, pp. 3029--3056, 2015.

\bibitem{samimi:2016:3-d-millimeter-wave-statistical-channel}
M.~K. Samimi and T.~S. Rappaport, ``{3-D} millimeter-wave statistical channel model for {5G} wireless system design,'' \emph{IEEE Transactions on Microwave Theory and Techniques}, vol.~64, no.~7, pp. 2207--2225, 2016.

\bibitem{samimi:2016:local-multipath-model}
M.~K. Samimi and T.~S.~Rappaport, ``Local multipath model parameters for generating {5G} millimeter-wave {3GPP}-like channel impulse response,'' in \emph{IEEE 10th European Conference on Antennas and Propagation (EuCAP)}, 2016, pp. 1--5.

\bibitem{qualcommVideo}
\BIBentryALTinterwordspacing
{Qualcomm Technologies, Inc.}, ``{VR} and {AR} pushing connectivity limits,'' Oct. 2018. [Online]. Available: \url{https://www.qualcomm.com/content/dam/qcomm-martech/dm-assets/documents/presentation_-_vr_and_ar_are_pushing_connectivity_limits_-web_0.pdf}
\BIBentrySTDinterwordspacing

\bibitem{tranter:2004:principles}
W.~Tranter \emph{et~al.}, \emph{{Principles of Communication Systems Simulation with Wireless Applications}}, 1st~ed.\hskip 1em plus 0.5em minus 0.4em\relax USA: Prentice Hall Press, 2003.

\bibitem{magrin:2019:simulation}
D.~Magrin \emph{et~al.}, ``{A simulation execution manager for ns-3: Encouraging reproducibility and simplifying statistical analysis of ns-3 simulations},'' in \emph{22nd International ACM Conference on Modeling, Analysis and Simulation of Wireless and Mobile Systems}, 2019, pp. 121--125.

\end{thebibliography}
